# Determination of conduction and valence band electronic structure of LaTiO$_x$N$_y$ thin film

Markus Pichler,[a] Jakub Szlachetko,[b],[c] Ivano E. Castelli,[d] Nicola Marzari,[d] Max Döbeli,[e] Alexander Wokaun,[a] Daniele Pergolesi *[a] and Thomas Lippert *[a],[f]

[a] M. Pichler, A. Wokaun, D. Pergolesi, T, Lippert  Laboratory for Multiscale Materials Experiments, Paul Scherrer Institut, 5232 Villigen-PSI, Switzerland
E-mail: daniele.pergolesi@psi.ch, thomas.lippert@psi.ch
[b] J. Szlachetko, Paul Scherrer Institut, 5232 Villigen-PSI, Switzerland
[c] J. Szlachetko, Institute of Physics, Jan Kochanowski University, Kielce, Poland
[d] I. E. Castelli, N. Marzari, Theory and Simulation of Materials and National Centre for Computational Design and Discovery of Novel Materials (MARVEL), École Polytechnique Fédérale de Lausanne, 1015 Lausanne, Switzerland
[e] M. Döbeli, Ion Beam Physics, ETH Zurich, 8093 Zurich, Switzerland.
[f] T. Lippert, Laboratory of Inorganic Chemistry, Department of Chemistry and Applied Biosciences, ETH Zurich, 8093 Zurich, Switzerland

**Abstract:** The nitrogen substitution into the oxygen sites of several oxide materials leads to a reduction of the band gap to the visible light energy range, which makes these oxynitride semiconductors potential photocatalysts for efficient solar water splitting. Oxynitrides typically show a different crystal structure compare to the pristine oxide material. Since the band gap is correlated to both the chemical composition and the crystal structure, it is not trivial to distinguish what modifications of the electronic structure induced by the nitrogen substitution are related to compositional and/or structural effects. Here, X-ray emission and absorption spectroscopy is used to investigate the electronic structures of orthorhombic perovskite LaTiO$_x$N$_y$ thin films in comparison with films of the pristine oxide LaTiO$_x$ with similar orthorhombic structure and cationic oxidation state. Experiment and theory show the expected upward shift in energy of the valence band maximum that reduces the band gap as a consequence of the nitrogen incorporation. But this study also shows that the conduction band minimum, typically considered almost unaffected by the nitrogen substitution, undergoes a significant downward shift in energy. For a rational design of oxynitride photocatalysts the observed changes of both the unoccupied and occupied electronic states have to be taken into account to justify the total band gap narrowing induced by the nitrogen incorporation.

**Introduction**

Oxynitrides are oxide where oxygen is partly substituted with nitrogen. The nitrogen incorporation into the native oxide alters the electronic properties making oxynitrides attractive as visible-light-driven photocatalysts,[1,2] dielectric materials[2] and non-toxic pigments.[3,4]

Especially in the field of solar water splitting, oxynitrides are considered as a particularly promising class of materials. The relatively large band gap of conventional oxide photocatalysts (typically ≥ 3 eV) limits the part of the solar spectrum that can be effectively utilized to the UV radiation (about 3%). The band gap of many oxynitrides is well below 3 eV shifting the photoresponse towards the visible light range, thus allowing a more efficient utilization of the solar radiation. Moreover, the bad gap energy of some oxynitrides encompasses both the reduction and oxidation potentials of water. This, together with band gap energy well below 3 eV, represents an ideal band structure for the simultaneous production of oxygen and hydrogen (overall water splitting) under solar light illumination.

Promising oxynitride materials are TaON with a band gap of about 2.5 eV,[5,6] GaN:ZnO solid solutions with a band gap of about 2.7 eV[7,8] and the perovskite-type $LaTiO_2N$ and $BaTaO_2N$ with band gaps of about 2.1 eV[9-11] and 1.8 eV,[12-16] respectively.

The narrowing of the band gap as a consequence of the nitrogen incorporation arises from the creation of additional electronic states just above the valence band due to the hybridization of N 2p orbitals at higher energy level with O 2p orbitals at lower energy level.[1,2] Thus, the valence band maximum (VBM) shifts upwards in energy, while the conduction band minimum (CBM) is considered to be almost unaffected by the nitrogen incorporation. This results in a reduced overall band gap.

Comparing for example $Ta_2O_5$, TaON, and $Ta_3N_5$ large upward energy shifts of the VBM were measured from the oxide to the nitride, while the CBM of the three materials was found similar.[17] However, the nitrogen incorporation in metal oxides like $Ta_2O_5$ and $TiO_2$ leads not only to a different chemical composition, but also to a different crystalline structure. Also in perovskite oxynitrides with the typical formula $ABO_xN_y$ the partial substitution of oxygen with nitrogen in the pristine oxide typically leads to a change of the crystalline structure. The VBM significantly shifts upward in energy depending on the N content reducing the band gap, while the width and the energetic position of the CB can be tuned by changing the A and the B cation respectively.[18,19] In particular, a different ionic radius of the A cation triggers a distortion of the $B(O,N)_6$ octahedra affecting the B-(O,N)-B bond angle. Moving from the cubic symmetry of $BaTaO_2N$ where the bond angle is 180° to the lower symmetries of $SrTaO_2N$ (tetragonal) and $CaTaO_2N$ (orthorhombic) where the bond angle is respectively 169.9° and 153.3°, the width of the CB decreases leaving the energetic center unaffected.[18,19] This results in a wider band gap. This effect is well-known for perovskite oxides. For instance, calculations on simulated $CaTiO_3$ perovskite structures with Ti-O-Ti bond angle ranging from 180° to 152° show that the width of the CB decreases (increasing the band gap) as the bond angle deviates from 180°.[20]

The comparison among Ba-, Sr-, and Ca-$TaO_2N$ provides an example of composition-induced structural modification that affects the band gap. Instead, in Ba- and Sr-$TaO_2N$ the substitution of the B cation $Ta^{5+}$ for $Nb^{5+}$, which have the same ionic radius of 0.88 Å in octahedral coordination, is an example of composition-induced modification of the band gap within the same crystal structure (cubic for Ba and tetragonal for Sr-based compounds). In perovskite oxides (and oxynitrides) the VBM is primarily set by the O (and N) 2p orbitals, while the CBM by the B cation transition metal d-orbitals. The valence-to-conduction band transition can be described as a charge transfer excitation from O (and N) to metal orbitals. It is thus expected that the electronegativity and the coordination environment of the B cation can affect the size of the band gap. These general considerations support the observation of a larger band gap for $(Ba,Sr)TaO_2N$ than for $(Ba,Sr)NbO_2N$, considering the lower electronegativity of Ta with respect to Nb.[18] However, as pointed out in reference (20) in general the electronegativity and coordination environment of the B cation are not the only parameters that must be taken into account, differently it would be difficult to explain the different band gap of $NaTaO_3$ and $KTaO_3$ or, considering binary oxides, the different band gap of anatase and rutile, or the smaller band gap of $WO_3$ as compared to $MoO_3$, and finally, considering oxynitride, the different band gap of $BaTaO_2N$ and $SrTaO_2N$ which originates from structural modifications as discussed above.

These considerations highlight the strong correlation between electronic and crystal structure. The band gap is sensitive to several compositional and structural factors unavoidably correlated making it difficult to isolate the influence of individual modifications of crystal structure and chemical composition.

For oxynitride materials, while the effect on the electronic structure of different A and B cations has been investigated quite extensively, to our knowledge the effects of the structural modifications related to the N substitution into a defined crystal structure are far from being completely characterized. This study aims at distinguishing these effects by comparing the electronic structure of oxide and oxynitride samples with the same crystalline structure and cationic composition. Achieving more insights into this topic may help understanding how photocatalytic materials can be modified to maximize their photo response in the visible light energy range.

For this study, the electronic structure of thin films of the perovskite oxynitride $LaTiO_xN_y$ (LTON) was determined by combining X-ray absorption and non-resonant emission spectroscopy (XAS and XES) and compared to isostructural lanthanum titanate (LTO) thin films.

The use of thin films is particularly suited for this investigation since by selecting different deposition parameters a careful experimental control on the structural properties and composition of the samples could be achieved.[21] LTON was selected due to its very promising properties. The band gap width and the energy position of the band edges make this material an almost ideal semiconductor for solar water splitting with the ability to generate both $H_2$ and $O_2$ under irradiation of light with wavelength up to ca. 600 nm. Moreover, the LTON composition is abundant, inexpensive and non-toxic.

**Results and Discussion**

LTON thin films, about 350 nm thick, were deposited on (0001)-oriented $Al_2O_3$ substrates by Pulsed Reactive Crossed-beam Laser Ablation (PRCLA) using ammonia for the reactive gas pulses and a target of $La_2Ti_2O_7$. Details on this modified pulsed laser deposition (PLD) method can be found elsewhere,[22,23] while the successful application of this technique for the growth of LTON films has been reported in literature.[21-24] For this work, the same experimental procedure reported in our previous study[21] was used.

Due to the different crystalline structure, the use of sapphire substrates is expected to lead to polycrystalline thin films. X-Ray Diffractometry (XRD) was performed at grazing incidence and the measured reflexes (Fig. 1a) confirm that the LTON films are isostructural to the stoichiometric LaTiO$_2$N orthorhombic perovskite (ICSD Coll. Code: 168551). Using Rutherford Backscattering (RBS) and Heavy-Ion Elastic Recoil Detection Analysis (ERDA), the composition was determined to be La$_{1.03}$Ti$_{0.97}$O$_{2.28}$N$_{0.66}$. In agreement with previous studies[21,30,31] the ablation process led to a slightly over-stoichiometric content of the heavier cation (La) at the expense of the lighter element (Ti). Slight shifts to smaller 2θ angles of the reflexes were measured compared to the stoichiometric composition of LaTiO$_2$N. This was already observed for LTON samples prepared by sputtering[32] and PRCLA[21] and can be explained by the smaller nitrogen content compared to the fully nitridated composition. Lower than stoichiometric nitrogen content is often observed also for powder samples due to the competing mechanisms of nitrogen incorporation into the LTON structure and the formation of secondary phases of TiO$_{1-x}$N$_x$ and TiN during prolonged high temperature ammonolysis. As a result of a nitrogen-to-oxygen ratio < 0.5 in oxynitride titanates, Ti is found in both the Ti$^{3+}$ and Ti$^{4+}$ oxidation states.

The same experimental set up was used for conventional PLD to grow the LTO thin films on the same kind of substrates. The phase diagram of the LaTiO$_x$ system has been investigated[33-35] in the range from LaTiO$_3$ to La$_2$Ti$_2$O$_7$ (3 ≤ x ≤ 3.5). In this system, completely different physical and structural properties depend on the oxygen content than controls the titanium valence between +4 and +3. The two extremes of the phase diagram (LaTi$^{3+}$O$_3$ and LaTi$^{4+}$O$_{3.5}$) are not suitable to be compared to LTON. On one side LTO with low oxygen content (x close to 3) has the same crystal structure as LTON (orthorhombic perovskite), however Ti is mainly in the Ti$^{3+}$ oxidation state (3d$^1$) and the material, depending on temperature and oxygen content, is a Mott-insulator or shows metallic-like electronic properties. On the contrary, at high oxygen content we have La$_2$Ti$_2$O$_7$ (x = 3.5), a semiconductor material where Ti is only in the Ti$^{4+}$ oxidation state (3d$^0$). Ti would be only in the 4+ oxidation state also in stoichiometric LaTiO$_2$N, which is not the case for our LTON films; nevertheless La$_2$Ti$_2$O$_7$ shows the monoclinic perovskite structure, which is different from the orthorhombic structure of LTON making it difficult to distinguish whether the modification of the electronic properties is primarily due to compositional or structural changes.

For the growth of suitable LaTiO$_x$ thin films N$_2$ was introduced as background gas during PLD with the intent to grow thin films with reduced oxygen content compared to La$_2$Ti$_2$O$_7$ (x < 3.5), thus with Ti in mixed Ti$^{3+}$ and Ti$^{4+}$ oxidation state (as in the LTON films fabricated by PRCLA), while still preserving the semiconducting nature of the material (x > 3.3)[33] but isostructural to the orthorhombic perovskite (x < 3.4)[33] of LTON.

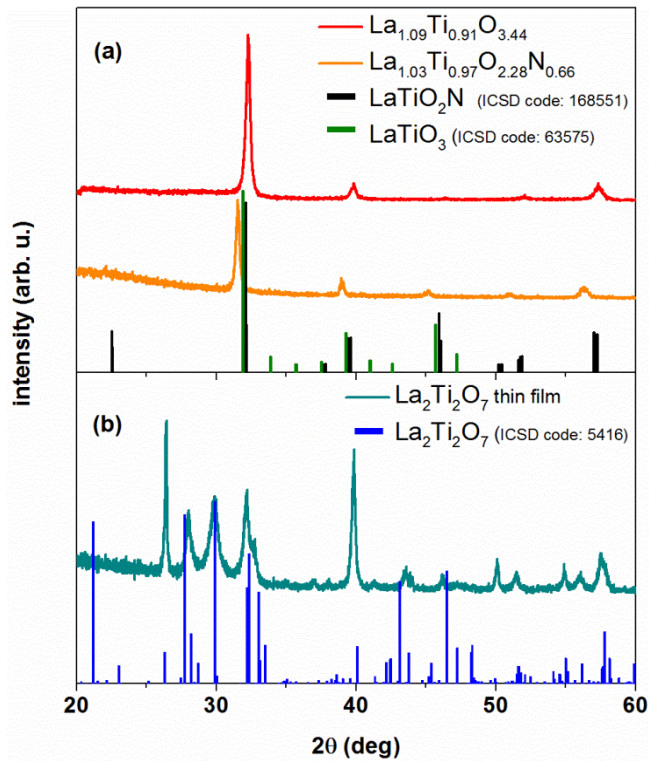

**Figure 1.** XRD analysis. (a) Grazing angle X-Ray diffraction pattern: (red) polycrystalline film of LaTiO$_{3+z}$ grown by PLD in N$_2$ environment; (orange) polycrystalline film of LaTiO$_x$N$_y$ grown by PRCLA using NH$_3$ for the gas pulse. The oxide and oxynitride films show the same orthorhombic crystal structure. (b) Grazing angle X-Ray diffraction pattern of a polycrystalline film of lanthanum titanate grown by PLD in oxygen environment showing the layered monoclininc crystal structure of La$_2$Ti$_2$O$_7$.

Fig. 1a shows the XRD pattern of the LTO film with reflexes at very similar 2θ values as observed for our LTON films indicating the orthorhombic perovskite structure. The angular positions of the reflexes are shifted toward slightly larger 2θ values compared to the stoichiometric composition of $LaTiO_3$ (ICSD Coll. Code: 63575), indicating a reduction of the unit cell volume. RBS and ERDA showed the film chemical composition to be $La_{1.09}Ti_{0.91}O_{3.44}$. According to the $LaTiO_x$ phase diagram proposed in literature[35] an oxygen content of 3.44 should lead to a monoclinic layered perovskite of the family $A_nB_nO_{3n+2}$ with n = 4.5, in contrast with the XRD analysis of the films that shows the orthorhombic structure. The growth of this material as thin film and/or the off-stoichiometric La/Ti ratio observed also for this film[30,31] may have helped the crystallization of the orthorhombic perovskite phase. Moreover, the ERDA measurement of the oxygen content has an uncertainty of ± 3%, thus considering the result of the XRD analysis it is likely possible that the measured oxygen content is actually overestimated. Indeed, for oxygen content between 3.29 and 3.40 high resolution electron microscopy revealed coexistence of the orthorhombic and monoclinic perovskite structures, while for oxygen contents above 3.40 only the layered monoclinic structure is allowed.[33,35] Fig. 1b shows the XRD analysis of a lanthanum titanate film grown for this work in oxygen environment with nominal composition $La_2Ti_2O_7$ (monoclinic crystal system). Fig. 1b also reports the angular position of the reflexes of the $La_2Ti_2O_7$ powder diffraction pattern (ICSD Coll. Code: 5416). The comparison of Fig. 1a and 1b clearly shows the orthorhombic phase of the LTO film grown in $N_2$ environment.

The LTON and LTO films not only show a similar crystal structure but the compositional analysis suggests that also the $Ti^{4+}$ and $Ti^{3+}$ contents are similar. X-ray photoelectron spectroscopy was used to probe the Ti 2p core level. As expected, the peak positions are consistent with the presence of Ti in both oxidation states. More importantly both, the Ti 2p peak positions and spectral shape are almost the same for LTO and LTON indicating very similar $Ti^{4+}$ and $Ti^{3+}$ contents (thus making these samples suitable for the purpose of this study). Hereinafter, we will refer to these two samples as LTON and LTO.

Fig. 2 shows the normalized transmittance of the two samples measured by UV-Vis spectrometry. LTO has a band gap of about 3.26 eV, i.e. transparent for the visible light range (between 390 nm to 700 nm), whereas a shift in the absorption edge for the oxynitride sample towards smaller photon energies of about 0.96 eV is observed. This confirms the effect of the incorporation of nitrogen on the reduction of the band gap. The estimated band gap value of LTON is about 2.30 eV. This value is about 0.2 to 0.3 eV larger than the literature values measured for LaTiO$_2$N powder samples,[9-11,36] which is related to the smaller nitrogen content of our films compared to the fully nitridated compound. Literature reports the photo-electrochemical characterization of LTON thin films grown on conductive substrates by PRCLA[21] and sputtering.[32] Good photo-electrochemical performances were measured depending on experimental conditions (electrolyte, co-catalyst loading) and crystalline quality of the LTON film.[32]

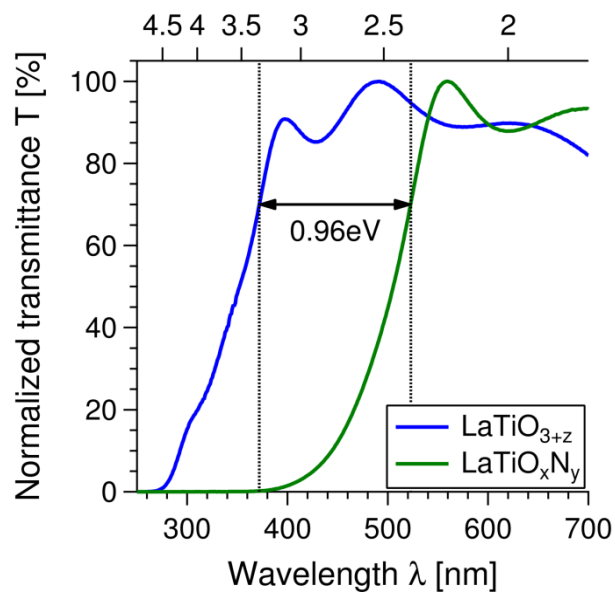

**Figure 2**. UV-Vis analysis. Normalized UV-Vis transmittance spectra of LaTiO$_{3+z}$ and LaTiO$_x$N$_y$ thin film samples grown on double-side polished Al$_2$O$_3$ substrates. Due to the thicknesses of the samples, the expected interference fringes above the transmittance edge are observed in the UV-Vis spectra. The dashed lines mark approximately the band gap values of the samples. For the ease of comparison, the transmittance data is normalized to 100%.

The electronic structure of the two films was probed by non-resonant XES and XAS.[37-48] XAS provides information about the unoccupied electronic states of an atom, whereas XES reflects the occupied electronic states, and when applied together, detailed information of the atomic orbitals can be achieved, giving an insight into the oxidation state, spin state, and chemical environment around a specific absorbing atom.

The investigation of the samples is performed according to the Szlachetko & Sá methodology for resonant XES data interpretation aiming at a rational design of photocatalytic materials.[41,49-52] The experimental XES and XAS data are compared with theoretical calculations of Density of States (DoS)[53,54] to assign the different features visible in the acquired spectra.

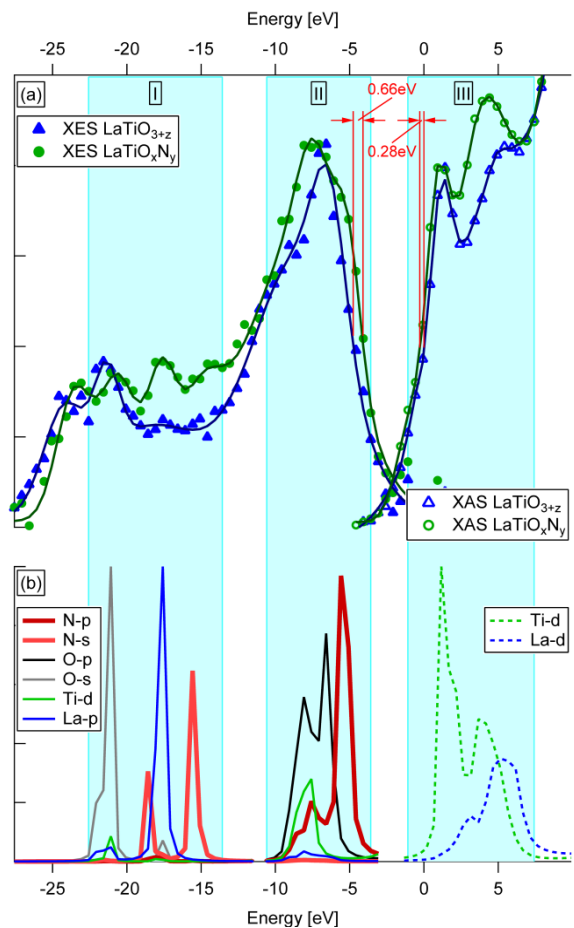

**Figure 3**. XAS and XES analysis. (a) Conduction and valence band electronic structure determined from X-ray emission and absorption spectroscopy measurements of $LaTiO_{3+z}$ and $LaTiO_xN_y$ thin films. (b) Calculated orbital contributions. (The markers in (a) correspond to the measured data and the lines act as a guide).

Fig. 3a shows the XES spectra (occupied states) representing the VB and the XAS spectra (unoccupied states) representing the CB for LTO (blue) and LTON (green). For both materials, the total XES and XAS spectrum is shifted in energy by 4967.5 eV, which is the energy value at the inflection point of the CB edge of the LTO film. This allows a better comparison of the measured data with the calculated orbital contributions. It is noteworthy that the measured XES and XAS spectra represent the p-projected DoS because the measurements are performed at Ti K-edge (i.e. s-type core-electron). The electronic transitions involved in the experiment are dominated by selection rules with highest probabilities for dipole transitions.

Fig. 3b shows the calculated orbital contributions (DoS).

Fig. 3 is further subdivided into 3 sections highlighted with light blue boxes:

- Section I (-22.5 eV to -13.5 eV) includes strongly bonded occupied states.
- Section II (-10.5 eV to -3.5 eV) includes the occupied states close to the VB edge.
- Section III (-1 eV to 7.5 eV) includes the unoccupied states.

In section I of the LTO XES spectrum, the two small features at about -17.5 eV and -21 eV can be identified with support of DoS calculations as La-p and O-s states, respectively. In the LTON spectrum of section I an additional feature at about -15 eV is detected and assigned to N-s states, while the feature at -17.5 eV became stronger and is composed of La-p and N-s states. On the other hand, the peak at -21 eV became smaller reflecting the reduction of the LTON oxygen content compared to LTO. Section II of the LTO XES spectrum shows the superposition of O-p and Ti-d states forming the distinctive peak at -6.5 eV. Comparing to the LTON XES spectrum, this peak shifts toward slightly lower energies while a relatively large shift toward higher energies of the VBM can be observed. The extent of this energy shift is about 0.66 eV, determined at the inflection point of the VBM of both samples. As mentioned above, according to literature the observed shift of the VBM to higher energies is explained with the lower electronegativity of the N 2p orbitals compared to O 2p orbitals.[2,49,55,56] A hybridization of the energetically higher N 2p and lower O 2p orbitals takes place, forming new electronic states just above the VB states of the native oxides thus shifting the VBM upwards in energy.

This explanation is confirmed by our results, where additional electronic states, assigned to N 2p states by theoretical calculations, can be measured experimentally with XES. Fig. 3b shows the calculated orbital contributions: the N-p orbitals form additional states at about -5.5 eV (about 1 eV higher in energy than the O 2p contribution).

However, the positive shift of 0.66 eV observed for the VB of LTON compared to LTO is not enough to explain the total band gap reduction of 0.96 eV (from 3.26 eV to 2.30 eV) observed in the UV-Vis transmittance spectra (Fig. 2). The additional energy levels created just above the VBM of LTO only account for less than 70% of the total band gap reduction. The analysis of the XAS spectra plotted in section III of Fig. 3a accounts for this inconsistency.

The XAS spectrum of LTO in section III consists of two features located at about 1 and 4 eV. The first is composed of Ti-d states, while the second arises from the hybridization of the Ti-d and La-d states (Fig. 3b).

The LTON XAS spectrum and theoretical simulations of DoS show that the oxygen to nitrogen substitution leads to no additional features in the unoccupied electronic states. However, a significant shift of the CBM of LTON to lower energies compared to LTO is visible in Fig. 3a. The extent of such a shift can be quantified in about 0.28 eV at the inflection point of the CBM of both samples. Together with the positive shift of 0.66 eV observed for the VBM, the total band gap reduction measured by XES and XAS is about 0.94 eV in very good agreement with the UV-Vis result.

Despite good agreement in band gap energy difference we should note that absolute energy values as extracted from X-ray spectroscopy measurement are approximately 1eV larger than those obtained from optical spectra. This effect is induced by the core-hole screening of 1s electronic state that is employed as a probe in the X-ray measurements. The influence of core-hole screening relates to sudden change of the electronic potential in many-body electron systems has been found relevant for both X-ray and electron XPS/Auger spectroscopies.[57-59] For data analysis we assumed that, independently on the extent of the core-hole screening effect to the spectra, the core-hole screening effect should be the same (i.e. introduce the same spectral effects) for both samples. To support this assumption, using the Bader analysis we calculated the amount of charges on each atoms for both samples. The electronic charge of the Ti ions in the lattice is equivalent to 20.01 and 19.95 electrons for LTO and LTON, respectively. This means that the charge on the Ti ions in the structure is basically the same for both compositions.

Since no additional orbital contributions arising from the nitrogen incorporation were detected in the unoccupied electronic states, the origin of the observed downward shift of the CBM is difficult to assign.

The shift of the CBM to lower energy may result from a stronger localization of the lowest unoccupied d-states of Ti due to the nitrogen incorporation. The effects of localized and delocalized orbitals on the energy position of the CBM were detected in the case of $TiO_2$ as additional pre-edge features visible in the XAS spectrum acquired in total fluorescence yield.[49,60]

As discussed above, in ternary oxides/oxynitrides $AB(O,N)_3$ the distortion of the unit cell, more specifically different values of the B-(O,N)-B bond angle, can affect the energy position of the CBM by reducing the width of the CB leaving the energetic center unchanged.[18-20] The lowest energy position of the CBM is found for the highest symmetry (cubic) with a B-(O,N)-B bond angle of 180°.

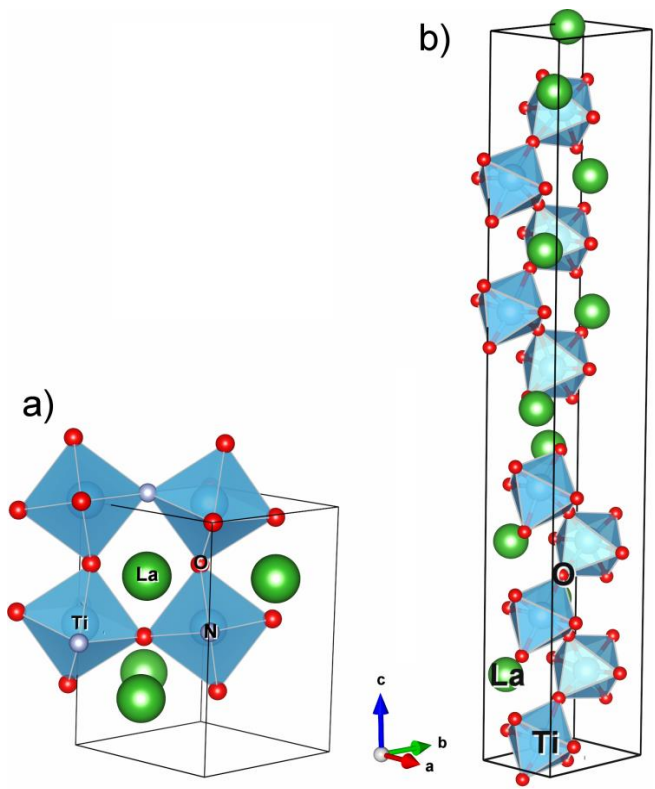

**Figure 4**. Simulations of the crystal structures of (a) $LaTiO_{2.25}N_{0.75}$ and (b) $La_5Ti_5O_{17}$ (i.e., $LaTiO_{3.4}$).

Lower symmetries with respect to cubic (tetragonal, orthorhombic) or stronger distortions within the same symmetry lead to larger tilt of the $B(O,N)_6$ octahedra, i.e. smaller B-(O,N)-B bond angle, thus raising up in energy the CBM. Simulations also show that within the same crystal structure of CaTiO$_3$ with same Pnma space group symmetry, same Ti-O and O-Ti-O bond distances, as the Ti-O-Ti bond angle deviates from 180° the band gap increases by shifting upward in energy the CBM.[20]

We speculate that a similar mechanism could explain our experimental results. Though in our case the LTON and LTO films both show the orthorhombic perovskite crystal structure, the N substitution could increase the width of the CB (lowering the CBM) by reducing the distortion of the unit cell. To support this hypothesis, we performed density functional theory simulations of the LaTiO$_{2.25}$N$_{0.75}$ and LaTiO$_{3.4}$ crystal structures in their orthorhombic symmetry (Fig. 4). The distortions of the octahedra was investigated using the GPAW code[62,63] and PBEsol as exchange-correlation functional.[64]

The calculations show that the Ti-(O,N)-Ti bond angle is about 161.5° in the orthorhombic LTON perovskite. For the oxide counterpart, the composition LaTiO$_{3.4}$ was selected because the oxygen content is close to that measured by ERDA and because, according to the LaTiO$_x$ phase diagram, x=3.4 represents the largest oxygen content allowed in the orthorhombic phase revealed by XRD for our samples. The idealized structure (not distorted) consist of a well-ordered stacking sequence of alternating layers of 5 unit cells along the c axis with a shift of (½ a, ½ b) between the layers. The simulation of the structure in the orthorhombic symmetry leads to a Ti-O-Ti bond angle around 158.0°. The result of the simulation and the comparison with literature[18-20] suggest that the N substitution in the semiconducting orthorhombic phase of LaTiO$_x$ (3.3 < x < 3.4) reduces the distortion of the unit cell which in turn lowers the band gap by widening the width of the CB.

It could be contended that the layered perovskite LaTiO$_{3.4}$ (belonging to the family A$_n$B$_n$O$_{3n+2}$ with n = 5) and the simple perovskite LaTiO$_{2.25}$N$_{0.75}$ (isostructural to LaTiO$_3$, i.e. n = ∞), though both in the orthorhombic crystal system, do not show the same crystal structure since the layering breaks the connectivity of the octahera along the c axis. This is indeed an important point in view of the discussed mutual relationship between the crystal and electronic structure. However, it has been shown that changing the composition of a simple perovskite to stabilize the layered phase causes only very small perturbations of the band gap.[20] In particular, the A$_{n+1}$Ti$_n$O$_{3n+1}$ (A = Sr, Ca) Ruddlesden-Popper phases were studied. Ca and Sr titanates have very similar band gaps for n = 2, 3, and ∞ the only exception being Sr$_2$TiO$_4$ (n = 1) that shows larger band gap than SrTiO$_3$. In this specific case, calculations suggest that about 0.2 eV larger band gap accounts for the distortion of the TiO6 octahedra of $Sr_2TiO_4$.[20] The authors generalized these findings concluding that reducing the three-dimensional octahedral connectivity of ternary perovskites (n = ∞) to a two-dimensional (n = 1) or pseudo two-dimensional (n > 1) connectivity does not impact directly the size of the band gap, while it does (widening the band gap) when the layering of the structure triggers a distortion of the octahedra.

By translating these results to the system La-Ti-O-N, we may assume that the higher CBM of $LaTiO_{3.4}$ with pseudo two-dimensional octahedra connectivity (5-layered stacking sequence) compared to the three-dimensional connectivity of the N-substituted $LaTiO_3$ arises (at least partly) from the increased distortion in the octahedra connectivity.

Beside the lattice distortion, also the electronegativity of the B cation influences the position of the CBM. Decreasing Increasing the electronegativity of the B cation causes an upward a downward shift in energy of the center of the CB without affecting the width, thus resulting in a wider smaller band gap[18]. We consider the fact that the lower CBM of LTON with respect to LTO could be caused by a change in the electronegativity in the anions with the consequent change of the charge transfer from the cations as well as of their valence. We calculate the charge transfer for LTO and LTON by means of the so-called Bader analysis.[65,66] For the LTO, in average La releases 2.20 electrons and Ti 2.0 1.99 electrons, while O receives around 1.2 electrons. For the LTON, La releases 2.10 electrons, Ti 2.05 electrons, while O and N receive 1.30 and 1.50 electrons, respectively.

Due to the lower electronegativity of N with respect to O, the charge transfer is indeed slightly different between the two materials. In particular, according to our calculation the charge transfer toward The different electronegativity of N and O, or a different hybridization between the metal and ligand states in the two materials, lead to a slightly different charge transfer. It is worth noticing that, according to our calculation the electronic charge transfer from the B cation is about 0.06 e larger for LTON than for LTO and this is qualitatively the same effect as that induced by a higher electronegativity of the B cation which leads to a downward shift of the CB. Moreover, the charge transfer from the A cation is about 0.10 eV smaller for LTON than for LTO, thus leading to a slightly larger average electronic charge on the cations in the case of LTON, as expected due to the overall lower electronegativity of the anions.

We conclude that, beside the expected upward shift in energy of the VBM, the N substitution into an isostructural physicochemical environment can also significantly affect the energy position of the CBM; in the specific case of the La-Ti-O-N system, the lower distortion of the octahedra in LTON compare to LTO enlarges the width of the CB leaving unaffected the energetic center. At the same time, the lower electronegativity of N with respect to O shifts Moreover, we suggest that also the lower electronegativity of N with respect to O may play a role in shifting the energetic center of the CB downward without affecting the width. Both effects can contribute to lower the CBM and thus the band gap.

**Conclusions**

In summary, lanthanum titanium oxynitride and lanthanum titanium oxide thin films were fabricated in their orthorhombic symmetry and with very similar cationic oxidation state to investigate the effect of nitrogen incorporation on the electronic and optical properties. Non-resonant X-ray emission and absorption spectroscopy were performed to map the occupied and unoccupied electronic states. Density of state calculations were used to assign the detected features in the X-ray spectra to atomic orbitals. As expected, the X-ray emission spectrum of the oxynitride clearly shows a shift of the valence band maximum to higher energies compare to the oxide due to the hybridization of O 2p orbital with the energetically higher N 2p orbital. However, the resulting band gap reduction does not match the value measured by UV-Vis spectroscopy. Only by taking into account the X-ray absorption measurement of a significant downward shift in energy of the conduction band minimum, normally considered to be unaffected by the nitrogen incorporation, the overall band gap reduction can be explained.

With the support of density functional theory simulations, we conclude that in perovskite oxynitrides the nitrogen incorporation can also significantly shift the energy position of the conduction band minimum compare to the oxide material by affecting the distortion of the unit cell and the anion-to-cation charge transfer.

**Experimental Section**

LaTiO$_x$N$_y$ thin films were deposited by a modified Pulsed Laser Deposition (PLD) method, namely Pulsed Reactive Crossed-beam Laser Ablation (PRCLA)[22-24]. This technique allows the nitrogen incorporation into the monoclinic oxide (La$_2$Ti$_2$O$_7$) by applying reactive and nitrogen-containing gas jets into the vacuum chamber (e.g. $NH_3$ or $N_2$) which cross the ablation plume close to the oxide target. Physicochemical interactions at the intersection of gas jet and plasma plume form new nitrogen containing species that are transferred to the substrate to form the oxynitride thin film. By adjusting the deposition parameters and the timing of the gas jets and laser pulses, $LaTiO_xN_y$ thin films can be produced[21,24]. For this work, (0001)-oriented $Al_2O_3$ substrates were used. Pt paste was used to provide good and homogeneous thermal contact between heater and substrate. The substrate temperature was set to 800°C and monitored with a pyrometer. The background pressure was increased from $5.0 \times 10^{-6}$ mbar to $8.0 \times 10^{-4}$ mbar using $N_2$. For the deposition, a sintered $La_2Ti_2O_7$ rod target and a $NH_3$ gas jet were used. The $NH_3$ gas jet was applied for 400 µs and are stopped 30 µs before the laser pulse hits the target. The laser fluence of the KrF excimer laser (λ = 248 nm, pulse width of 30 ns) was adjusted to 3.9 J cm$^{-2}$ and a repetition rate of 10 Hz was used. During the deposition, the measured background pressure increased to $1.0 \times 10^{-3}$ mbar. The target-to-substrate distance was adjusted to 50 mm. Conventional PLD was used to grow the $LaTiO_x$ samples using the same $La_2Ti_2O_7$ rod target and the same kind of substrate. The substrate temperature was set to 800°C. $N_2$ was introduced as background gas with a partial pressure of 0.1 mbar to grow a thin film of $LaTiO_x$ with reduced oxygen content compared to the $La_2Ti_2O_7$ target material. A repetition rate of 10 Hz and a laser fluence of 2.0 J cm$^{-2}$ were applied. The target-to-substrate distance was fixed to 45 mm.

The film thickness was determined by surface profilometry (Dektak 8, Veeco Instruments). A Bruker-Siemens D500 X-Ray Diffractometer was used to determine the crystalline properties of the samples and a Cary 500 Scan UV-Vis-NIR spectrophotometer was used to measure the transmittance in the wavelength range from 200 nm to 1000 nm (6.20 eV to 1.24 eV). Rutherford Backscattering (RBS) and Heavy-Ion Elastic Recoil Detection Analysis (ERDA) were performed to investigate the chemical composition. RBS uses a 2 MeV $^4$He beam. The backscattered $^4$He particles were detected by a silicon PIN diode detector under an angle of 168° towards the incident beam. The RUMP program was used for the analysis. For ERDA, the samples were irradiated with a 13 MeV $^{127}$I beam. The recoiled particles were detected via a time-of-flight spectrometer combined with a gas ionization detector.

The X-ray Absorption Spectroscopy (XAS) and non-resonant X-ray Emission Spectroscopy (XES) experiments were performed at the SuperXAS beamline of the

Swiss Light Source, Switzerland. The X-rays, delivered by bending magnets, were monochromatized with double Si(111) crystals, and focused down to a size of 100 x 100 µm$^2$ by a Pt mirror. The X-ray absorption spectra were measured around the Ti K absorption edge at 4966 eV in total fluorescence mode by a Silicon Drift Detector. A fixed incidence X-ray energy at 5000 eV was used for measurements of the X-ray emission spectra for Ti valence-to-core transitions and by a dispersive-type von Hamos spectrometer[25] at curvature radius of 25 cm. The spectrometer was aligned in vertical scattering geometry with the Ge(400) reflection and the Bragg angle around 62°. To preserve high energy resolution for X-ray detection, we employed a segmented-type crystal consisting of 50 x 1 mm$^2$ segments, with a total crystal dimension of 50 x 100 mm$^2$ (dispersion x focusing). The X-rays diffracted by the crystal, were registered with a 2D Pilatus detector having a pixel size of 175 µm. The sample was aligned in the grazing incidence geometry, in order to enhance the XES signal recorded by the von Hamos spectrometer. We would like to note that the so-obtained line-source does not influence the overall energy resolution of the setup, but greatly increase number of events registered by the spectrometer.[26]

Theoretical simulations of the Density of States were performed with FEFF9.0 code[27] commonly used for X-ray spectroscopic data interpretation. FEFF9.0 is a self-consistent multiple-scattering code able to simulate electronic structure of material based on structural information.[28,29] The simulations retrieve the electronic orbital contribution for occupied and unoccupied states. In the computations we assumed for the N-doped sample the substitutional replacement of O to N atoms. We should also note that due to the 1s core-state probed in the experiment the measured X-ray spectra reflects p-projected density of states and therefore, the intensities of the measured spectral features will differ from the calculated density of states contributions.

**Acknowledgements**

This study received support from Swiss National Science Foundation (SNSF) grant agreement No IZERZ0_142176, from the National Centre for Competence in Research Discovery of new Materials (MARVEL) and for the Paul Scherrer Institut. The authors would like to acknowledge for the access to the SuperXAS beamline at Swiss Light Source, Switzerland, and the help from its team. J.S. acknowledges National Science Centre, Poland (NCN) for support under grant no.

2015/18/E/ST3/00444. I.E.C thanks EPFL Fellows co-funded by Marie Sklodowska-Curie (CE-COFUND Fellow - Fund 587704) for support. The authors also gratefully acknowledge Prof. Christophe Copéret for fruitful discussion.

**Keywords:** oxynitrides • Solar water splitting • X-ray absorption and emission spectroscopy • Electronic structure • Pulsed laser deposition